%% file: paper.tex
\documentclass[conference]{IEEEtran}

\usepackage{epsfig}
\usepackage{multirow}
\usepackage{microtype} 
\usepackage{booktabs}  
\usepackage{url}
\usepackage{paralist}
\usepackage{subfig}
\usepackage{todonotes}
\usepackage{amsmath}
\usepackage{algpseudocode}
\usepackage{algorithm}
\usepackage{algorithmicx}
\usepackage{url}

\algnewcommand{\algorithmicgoto}{\textbf{go to}}%
\algnewcommand{\Goto}[1]{\algorithmicgoto~\ref{#1}}%
\algrenewcommand\algorithmicindent{1.0em}%
\algnewcommand{\LineComment}[1]{\State \(\triangleright\) #1}

\usepackage{color}

\begin{document}

\date{}

\title{Design Automation for Obfuscated Circuits with Multiple Viable Functions}
\author{\IEEEauthorblockN{Shahrzad Keshavarz\IEEEauthorrefmark{1}, Christof Paar\IEEEauthorrefmark{2},
and Daniel Holcomb\IEEEauthorrefmark{1}
\IEEEauthorblockA{\IEEEauthorrefmark{1}ECE Department, University of Massachusetts Amherst,
Amherst, MA, 01003, USA
\\ \{skeshavarz, dholcomb\}@umass.edu 
}
\IEEEauthorblockA{\IEEEauthorrefmark{2}Horst G{\"o}rtz Institute for IT Security,  Ruhr-Universit{\"a}t Bochum, Bochum, Germany\\{christof.paar@rub.de}
}}
}
\maketitle

\begin{abstract}
\input{abstract}
\end{abstract}

\section{Introduction}
\label{sec:introduction}
\input{introduction}


\section{Setting}
\label{sec:setting}

\input{setting}

\section{Problem Formulation}
\label{sec:approach}
\input{approach}

\input{newalg}

\section{Evaluation}
\label{sec:evaluation}
\input{evaluation}

\section{Conclusion}
\label{sec:conclusion}
\input{conclusion}

\thispagestyle{empty}


\bibliographystyle{acm}
\bibliography{refs}

\end{document}

%% file: abstract.tex
Gate camouflaging is a technique for obfuscating the function of a circuit against reverse engineering attacks. However, if an adversary has pre-existing knowledge about the set of functions that are viable for an application, random camouflaging of gates will not obfuscate the function well. In this case, the adversary can target their search, and only needs to decide whether each of the viable functions could be implemented by the circuit.

In this work, we propose a method for using camouflaged cells to obfuscate a design that has a known set of viable functions. The circuit produced by this method ensures that an adversary will not be able to rule out any viable functions unless she is able to uncover the gate functions of the camouflaged cells. Our method comprises iterated synthesis within an overall optimization loop to combine the viable functions, followed by technology mapping to deploy camouflaged cells while maintaining the plausibility of all viable functions. We evaluate our technique on cryptographic S-box functions and show that, relative to a baseline approach, it achieves up to 38\% area reduction in PRESENT-style S-Boxes and 48\% in DES S-boxes. 


%% file: introduction.tex
The are many reasons that a chip designer may wish to prevent a reverse engineer from learning the exact function implemented on a target chip. He may, for example, want to prevent IP theft or prevent an adversary from learning information about the architecture which would allow her to mount side channel attacks.

Partially prompted by the increasing practicality of invasive reverse engineering attacks, there has been recently several proposals for using look-alike cells to obfuscate circuit functions, sometimes coined gate camouflage. For example, works have proposed to do this using oxide-terminated vias~\cite{chow-2007-integrated,rajendran-13}, using different threshold voltages~\cite{iyengar-15,collantes-16}, or using doping to create stuck-at-faults~\cite{becker-13}.
In the scenario where intermediate values can be read through a scan chain, SAT-based attacks are applicable~\cite{DuoLiu,elmassad-15}.
\par
Given that an adversary won't know the exact function of each look-alike cell, she must consider an exponential set of {\em plausible} functions that the circuit may implement. We use the terminology plausible function of a circuit to denote a function that a circuit or sub-circuit could implement given its use of camouflaged cells. Starting with a synthesized circuit, a designer replaces ordinary cells with camouflaged look-alike cells, and in doing so implicitly creates the exponential set of plausible functions that is guaranteed to contain the true function as well as many other (incorrect) functions. Yet, even a set with exponentially many plausible functions may not fool an attacker who knows that only specific functions are {\em viable} for the chip's application.  For instance, for an obfuscated arithmetic function, it is usually easy for an attacker to extract the correct functionality.
An attacker can check whether the plausible functions contain a particular function of interest (e.g. a viable function) by checking satisfiability of a QBF problem that is similar to equivalence checking, but with unconstrained side inputs that select which of the plausible functions is realized by the circuit~\cite{subramanyan-14}.
Previous works have not addressed how to obfuscate against such an adversary with knowledge of viable functions, and have implicitly assumed that the attacker sees all plausible circuit functions as viable ones.

In this work, we consider how to obfuscate a circuit against an adversary that has prior knowledge of a fixed set of viable functions that might be implemented in an obfuscated design.
In the setting that we consider, an adversary is not able to query inner circuit values directly, and is seeking to reverse engineer the logic function of a circuit using the following capabilities:

\begin{itemize}
	
	\item She has knowledge of the cell library used in the design, including camouflaged look-alike cells. 
	
	\item She can identify the cells and their connections by imaging the delayered circuit and matching against the components from the library. Her knowledge of the library allows her to infer the plausible functions of each look-alike cell instance, but she does not specifically know which of the functions is implemented by each cell instance.
	
	\item She has pre-existing knowledge of a specific set of viable logic functions $F = (f_1,\dots,f_n)$ for the circuit.
	
\end{itemize}



The number of different n-input Boolean functions is doubly exponential in the number of inputs, whereas the number of plausible functions in a camouflaged combinational circuit is, at most, exponential in the number of camouflaged cells. Therefore, it is improbable that the viable functions will be found in the set of plausible functions unless they are intentionally made to be plausible. This implies that random camouflaging is insufficient for obfuscating viable functions. In this paper we go beyond random camouflaging to present an automation strategy that the designer can use to achieve his goal of making plausible all of the viable functions.
\par
The specific contributions made in this paper are as follows.
\begin{itemize}
	
	\item A description of the obfuscation problem in which the adversary has partial knowledge about circuit function but lacks ability to query the direct outputs of the circuit.
	
	\item A novel design automation strategy using synthesis, heuristic optimization, and technology mapping to obfuscate circuits in a way that makes a set of chosen functions all appear plausible. 
	
	\item Evaluation of the approach on cryptographic S-box circuits, showing an area reduction of up to 48\% in DES S-boxes and up to 38\% in PRESENT S-boxes~\cite{bogdanov-07} compared to an approach that does not employ our method.
	
\end{itemize}

%% file: setting.tex
Although our technique is compatible with any library of camouflaged cells, we use cells that are constructed by modifying the doping of nominal library cells. By modifying doping to turn transistors ON and OFF, a cell can be made to implement the positive and negative co-factors of its nominal function with respect to each input.

\begin{figure}
	\centering
	\subfloat[An example 2-input NAND gate]{
		\includegraphics[width=0.22\columnwidth]{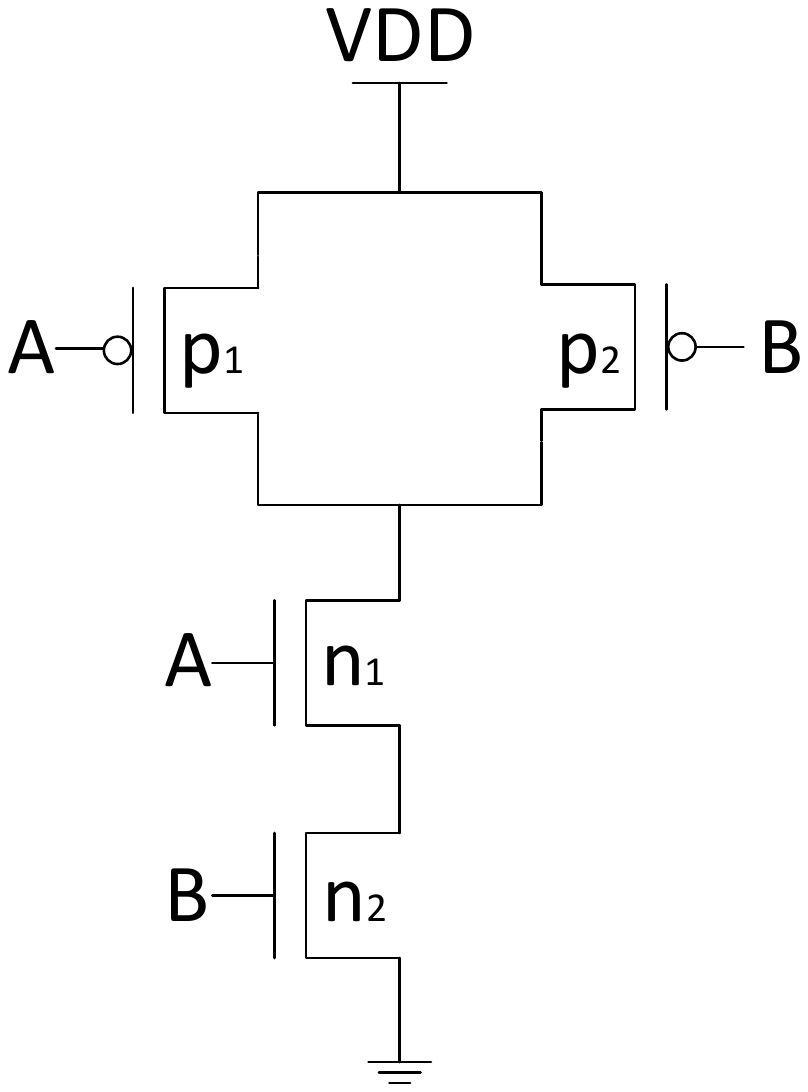}
		\label{fig:nand2}
		}
		\hspace{1cm}
			\subfloat[The truth table of functions that can be realized by a 2-input NAND gate with non-standard doping]{
				\includegraphics[width=0.37\columnwidth]{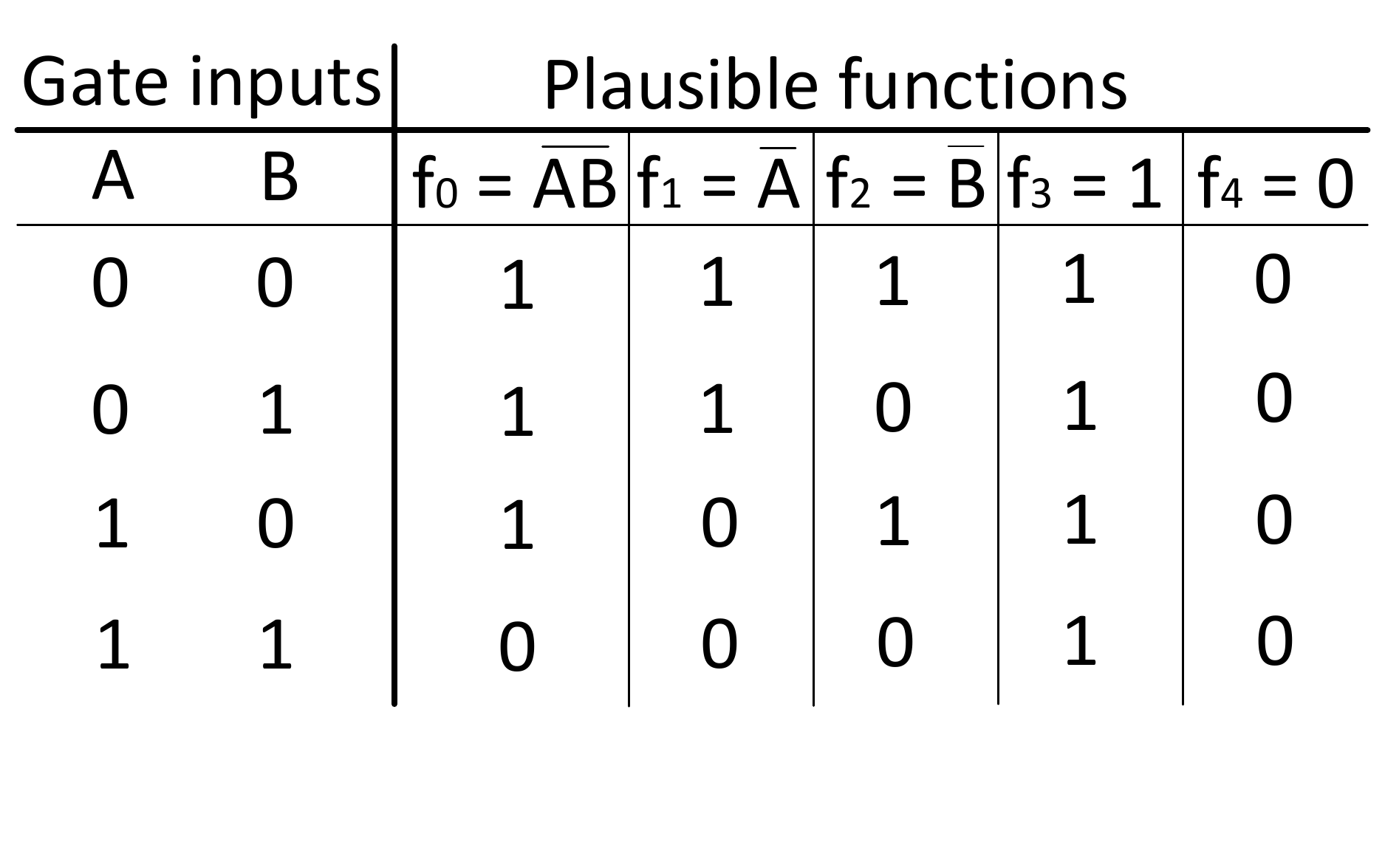}
				\label{fig:TT}
			}
	\caption{The library cell for a 2-input NAND gate}
	\label{fig:LibraryCell}
	\vspace{-10pt}
\end{figure}

For example, consider the camouflaged 2-input NAND as shown in Fig.~\ref{fig:nand2}. The nominal function of the cell is $f=\overline{AB}$. In a variant where $p_2$ is always ON and $n_2$ is always OFF, the cell implements $f_{\overline{B}}$, which is constant '1'. On the other hand, if $p_2$ is always OFF and $n_2$ is always ON, then the cell is implementing $f_B$ which is $\overline{A}$. The cell can be similarly modified to implement $f_{\overline{A}} = 1$, $f_A = \overline{B}$, and $f_{\overline{A}\overline{B}} = 0$. Fig.~\ref{fig:TT} shows the truth table of all possible functions that can be achieved by changing the transistor doping of a 2-input NAND gate. We use the same approach to create camouflaged versions of the other library cells as well.

%% file: approach.tex
\textit{}
We propose a three-phase approach for synthesizing circuits that can plausibly implement a chosen set of functions. We use logic synthesis to create a merged logic circuit with the capability to implement all the functions. We then add heuristic optimization to decide how to impose the functions onto each other in a way that maximizes logic sharing in synthesis. Lastly we perform technology mapping to cover the synthesized circuit using the plausible functions of the camouflaged look-alike cells, to reduce cost while preserving security. 


\subsection{Phase I: Multi-Function Synthesis}
\begin{figure}[htb!]
	\centering	 \includegraphics[width=0.82\columnwidth]{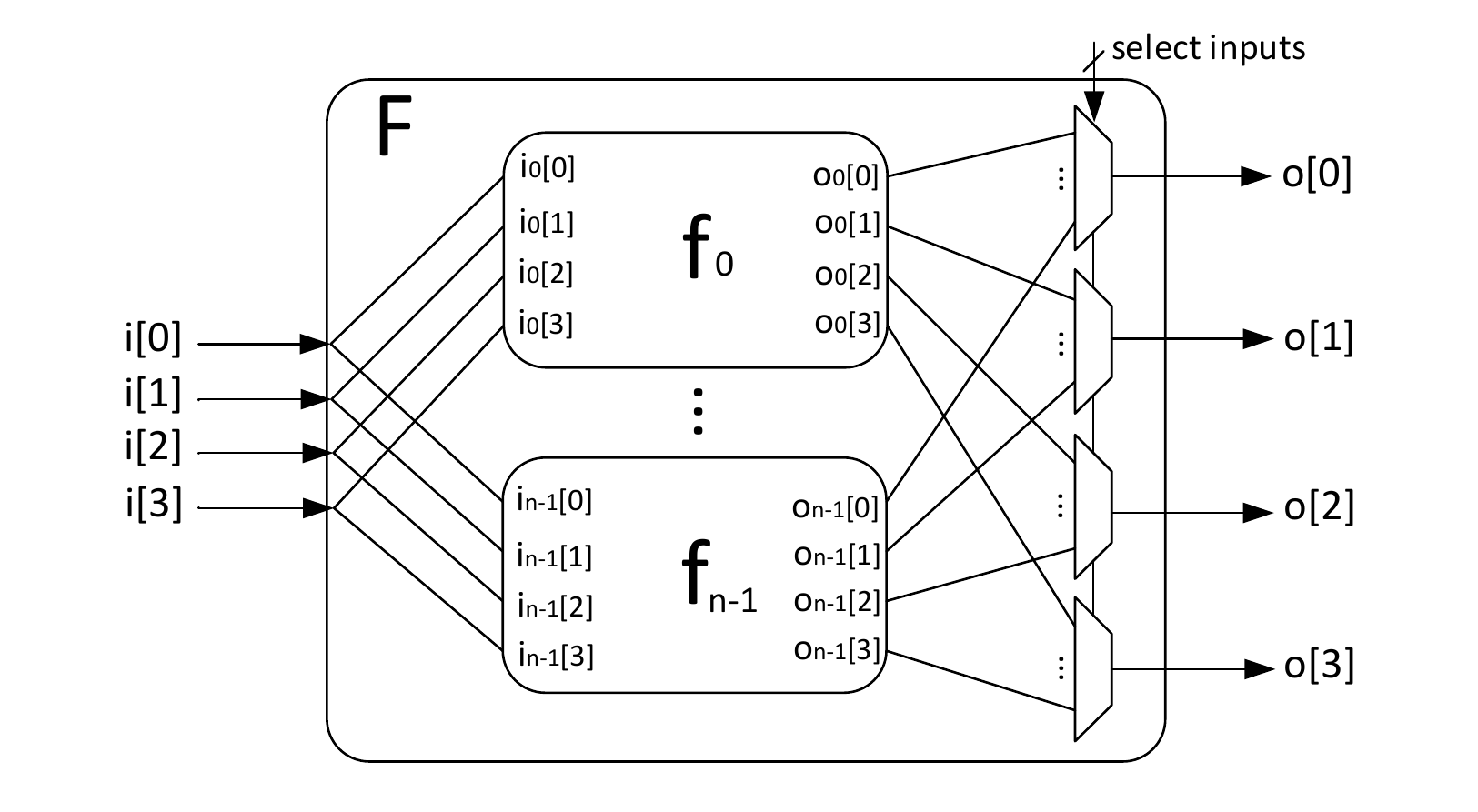}
	\vspace{-5pt}
	\caption{High level schematic for the circuit merging $n$ different 4-input 4-output functions, such as 4-bit S-boxes}
	\label{fig:HighLevelCircuit}
\end{figure}

To have a general circuit that can implement viable functions $(f_0, f_1, \dots, f_{n-1})$, we write RTL for a design that contains all of the functions with shared input signals, and add multiplexers at the outputs to choose between the outputs of the different functions. Fig.~\ref{fig:HighLevelCircuit} shows the high level schematic of this merged circuit for $n$ functions, each with 4 inputs and 4 outputs. The select inputs to the multiplexers choose which function's output will be the overall output of the circuit, and therefore, for appropriate assignment to the select inputs, the merged circuit is equivalent to any of the viable functions. 

The merged design containing all viable functions is then synthesized to produce a gate-level design where the select signals are inputs that may be used anywhere in the circuit, and not only right at the outputs. We use ABC~\cite{ABC} for synthesis to enhance logic sharing and minimize area with our own script comprising multiple $refactor$, $rewrite$ and $balance$ commands. Because ABC has limited input syntax, we use Yosys~\cite{yosys} to map RTL into a blif netlist that can be read by ABC. ABC maps the blif to a set of logic gates comprising inverters, buffers, and 2-4 input NAND, NOR, AND, OR gates.




\subsection{Phase II: Maximizing Logic Sharing}\label{ss:num2}
Circuit synthesis of a merged design will inherently try to share logic across the viable functions in order to minimize area. However, the potential for logic sharing depends on the input and output pin assignments of the merged functions. Assuming that an adversary doesn't know which specific signals are carried on particular input or output wires, she must consider a function to be plausible as long as there is some input and output interpretation that causes the obfuscated block to plausibly implement that function. The designer can exploit this degree of freedom to choose the input and output correspondence across the viable functions in a way that will maximize logic sharing.



\begin{figure}[htb!]
	\centering
	\subfloat[effective input placement for sharing]{	
		\label{fig:goodSharing}
		\includegraphics[width=0.95\columnwidth]{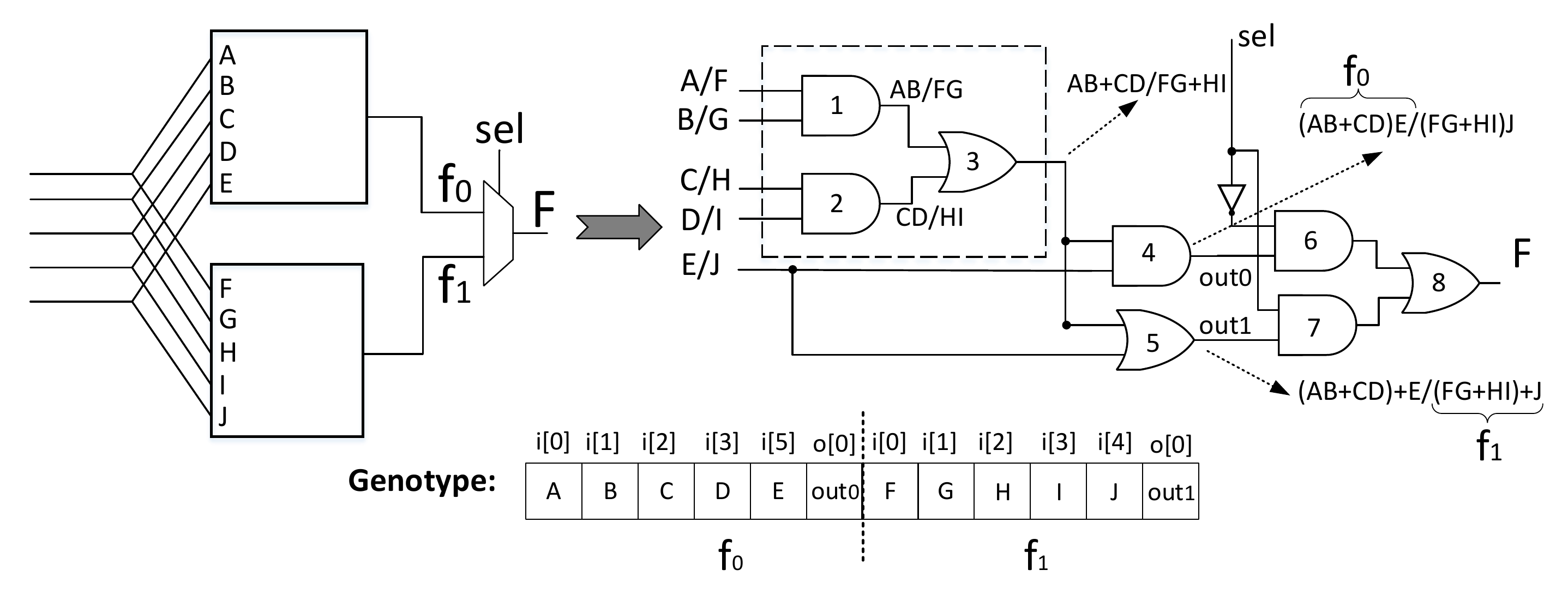}

	}
        \newline 
        \centering
	\subfloat[ineffective input placement that allows less sharing]{
		\includegraphics[width=0.95\columnwidth]{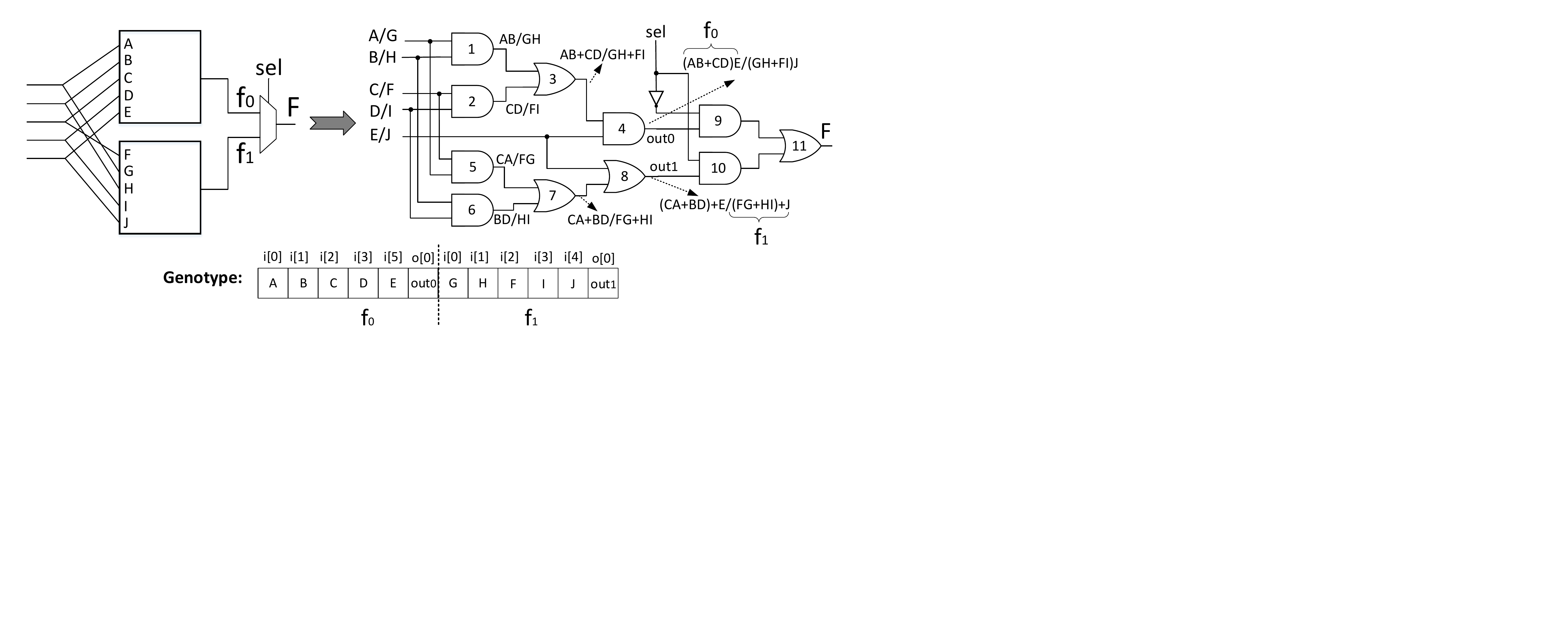}
		\label{fig:badSharing}
	}
	\caption{An example showing the importance of input positioning for logic sharing}
	\label{fig:positioning}
\end{figure}

Fig.~\ref{fig:positioning} shows two different ways of mapping the functions $f_0 = (AB+CD)E$ and $f_1=(FG+HI)+J$ onto each other. A designer that wants to show both functions as plausible must decide which input of $f_0$ corresponds to each input of function $f_1$. The mapping in Fig.~\ref{fig:goodSharing} is preferable because it allows the sub-circuit surrounded by a dotted line to be used in both functions $f_0$ and $f_1$. However, in Fig.~\ref{fig:badSharing}, the input placement does not allow the same extent of sub-circuit sharing between functions $f_0$ and $f_1$, and more gates are needed to implement the function. This example shows that assigning input position of each function can increase opportunities for logic sharing between functions and can hence reduce redundant logic to save area. The same observation about effective pin assignment also holds true for outputs when the respective functions have multiple outputs.

When the number of viable functions is large, it is infeasible to find the best pin assignment by exhaustive search. Furthermore, random search may not yield a good solution. To address this issue, we find effective pin assignments using genetic algorithm with the Python Package DEAP~\cite{deap}. The fitness function used to evaluate the quality of a pin assignment is the synthesized circuit area as reported by ABC. Therefore, we are using repeated logic synthesis in our exploration of pin assignments to try to find a pin assignment that will minimize area by enabling a high degree of logic sharing across the functions.

The genotype of genetic algorithm is a vector that specifies the pin assignments of the viable functions. For inputs, the genotype specifies which input pin of each viable function will share the same input pin of the overall merged circuit. For outputs, the genotype specifies which output pins of each viable function will connect to the function-selecting multiplexer of each output in the merged circuit. The fitness of genotypes are evaluated using the area reported by synthesis as explained above. Area is a useful objective because it encourages configurations that maximize sharing.Genotype instances with high fitness (low area) are propagated using mutation and crossover.
\begin{figure}[htb!]
	\centering
	\captionsetup[subfloat]{farskip=-10pt,captionskip=-10pt}
	\subfloat [Area distribution when using random input pin assignments]{
		\includegraphics[width=0.85\columnwidth]{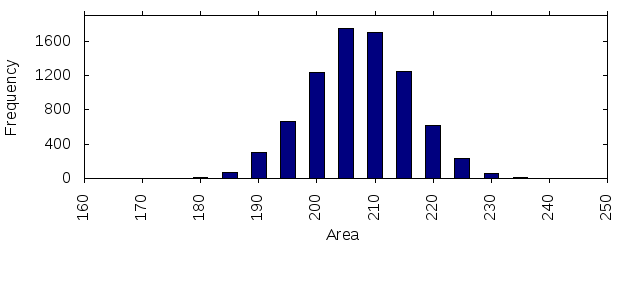}
		
		\label{fig:histogram}
	}\newline
	\subfloat [Area from genetic algorithm surpasses best random]{
		\label{fig:genVSrand}
		\includegraphics[width=0.85\columnwidth]{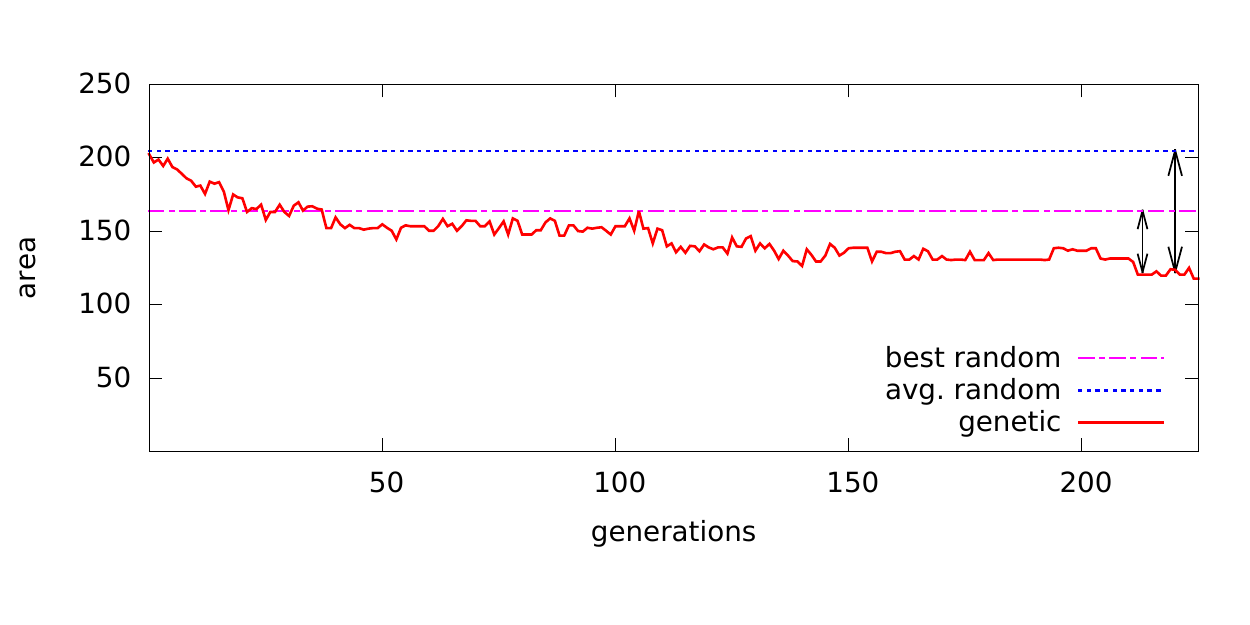}	
	}
	\caption{Synthesized circuit area of 8 merged PRESENT S-boxes when pin assignment is random or chosen by genetic algorithm.}
	\label{fig:area}
\end{figure}

Figures~\ref{fig:histogram} and~\ref{fig:genVSrand} together show that the genetic algorithm is able to find solutions that use less area than what can be achieved by trying random configurations. Fig.~\ref{fig:genVSrand} shows how the area improves across iterations of genetic algorithm. The reported areas are in units of GE (gate equivalents) which is the circuit area normalized to the area of a NAND2 gate in the same technology. The x-axis shows the number of generations in genetic algorithm, with each generation creating and evaluating a number of individuals. For comparison, we also evaluate a number (9726) of random pin assignments that is equal to the number of individuals evaluated during the genetic algorithm process; the distribution of areas from the random individuals is shown in Fig.~\ref{fig:histogram}. The area of the average and best random solutions are drawn as horizontal lines on Fig.~\ref{fig:genVSrand} to show visually that the genetic algorithm method is clearly finding pin assignment solutions that surpass what can be achieved by generating the same number of configurations randomly.


\subsection{Phase III: Technology Mapping to Deploy Cells}\label{ss:num3}

The synthesized merged circuit has a number of logical ``select'' inputs that choose between the viable functions. This circuit gets mapped to a circuit with camouflaged gates such that all viable functions in the synthesized circuit become plausible functions in the mapped circuit. One could accomplish this by adding a stealthy mechanism to connect each select signal to supply or ground; the attacker, without knowing the values of the select signals, would not be able to rule out any viable functions. However, instead of assigning values to the select signals, we use technology mapping to reduce the area cost of the circuit and eliminate the select signals. As will be explained, the key to this mapping is ensuring that, locally for any subcircuit with camouflaged cells, the plausible functions of the subcircuit include all corresponding functions of the synthesized circuit for any assignment to its select inputs. Meeting this condition ensures that all viable functions that were plausible in the synthesized circuit will remain plausible in the mapped circuit. 

As is common in technology mapping~\cite{keutzer1987dagon}, our approach decomposes the circuit graph into trees and uses dynamic programming tree covering to map the trees into cells. Each tree describes a fanout-free subcircuit with a single output that implements some Boolean function over the leaf nodes of the tree. The significant difference between our approach and ordinary technology mapping by tree covering is that in ordinary technology mapping, a subtree can be mapped to a cell if the cell's single function is {\itshape equivalent to} the subtree's single function. In our approach, since we have to consider multiple functions depending on the value of the select signals, a subtree can be mapped to a cell if the cell's plausible functions {\itshape contain} all of the desired functions for the subtree.



To allow tree covering to be used on the synthesized circuit, we first create a forest of trees from the circuit by splitting it at all fanouts. We then use the tree-covering procedure described in Alg.~\ref{alg:tmap} for mapping each tree to camouflaged cells. The algorithm uses dynamic programming starting from the leaf nodes of the tree and working toward the output. Whenever a node is considered for mapping, minimum-cost mappings will have already been discovered for all nodes in its transitive fanin. The function \Call{absfunc}{} in line~\ref{algline:ABSfunc} abstracts away any select signals in the tree; in doing so the logic function of the tree is translated into the set of logic functions that it can implement for any assignment to the select signals in the tree. To cover a node, different-sized subtrees having that node as output are candidates to be mapped to a new camouflaged cell. Each candidate subtree will have different leaf nodes from the other candidate subtrees. If a subtree is mapped to a cell, the cost of that covering is the cost of that cell added to the cost of of the optimal coverings of all of its leaf nodes (line~\ref{algline:cost} of Alg.~\ref{alg:tmap}). The lowest cost covering is chosen for each node in the tree until the entire tree is covered.

%% file: newalg.tex
\small
\begin{algorithm} 
\scriptsize
\caption{\footnotesize Technology mapping to cover a tree with camouflage cells to eliminate the select inputs while preserving as plausible all functions of the output node that could occur under any assignment of the select inputs.}
  \begin{algorithmic}[1]	
    \Function{tree-cover}{t} 
    \State{$cost(n_i) \gets \infty \quad \forall$ nodes $n_i \in t$}
    \ForAll{node $n_i \in t$, in topological order}
    \ForAll{subtree $t_s$ with output $n_i$ and depth $<3$}
    \LineComment{leaves of $t_s$ are already-covered nodes in $t$}
    \State{$F(t_s) \gets $\Call{absfunc}{}$(t_s)$}\Comment{functions to preserve} \label{algline:ABSfunc}
    \ForAll{camouflage library cell $g_j$}
    \If{plausiblefunctions($g_j$) $\supseteq F(t_s)$}\label{algline:cover}
    \LineComment{cell $g_j$ contains all functions of $t_s$}
    \State{$c \gets cost(g_j) + \sum_{n_k \in Leaves(t_s)} cost(n_k)$}\label{algline:cost}
    \If{$c < cost(n_i)$} 
    \State{$cost(n_i) \gets c$} \Comment{new opt. cover for $n_i$}

    \LineComment{\parbox[t]{2.35in}{Cover $n_i$ by mapping $t_s$ to cell $g_j$ and using optimal covers for leaf nodes of $t_s$}}
    \EndIf
    \EndIf
    \EndFor
    \EndFor
    \EndFor
    \State{\Return mapped circuit for tree} 
    \EndFunction
  \end{algorithmic}
\label{alg:tmap}
\end{algorithm}
\normalsize
\setlength{\textfloatsep}{3pt}

%% file: evaluation.tex
\normalsize
To evaluate our proposed method, we use the 16 different 4-bit S-box functions from Leander and Poschmann~\cite{leander-sbox} as viable functions. Each such S-box is a 4-bit-input 4-bit-output function that requires around 30 gate equivalents to implement. We create obfuscated designs that plausibly implement 2, 4, 8 or all 16 of the S-box functions in a single circuit. Additionally, we create obfuscated designs that plausibly implement 2,4, or all 8 of the  6-bit-input 4-bit-output DES S-boxes (each of these S-boxes are around 150 gate equivalents in area). We use genetic algorithm as discussed in \ref{ss:num2} and generate random pin position assignments equal to the number of total individuals that are evaluated in genetic algorithm. We then use the technology mapping algorithm from \ref{ss:num3} to map the resulting  circuits into our camouflaged library cells. To validate the correctness of our implementation, we verify using ModelSim that the resulting circuits can implement each of the viable functions when appropriate gate functions are supplied. Tab.~\ref{table:table1} reports the synthesized area for the best case and average case of random pin assignment, as well as the area when genetic algorithm is used (GA), and the area when genetic algorithm is followed by technology mapping (GA+TM); all areas are given in units of GE (gate equivalents). 

As can be seen, when comparing our final area to the synthesized result for the best randomly discovered pin assignment, our techniques provide an area improvement of up to 38\% for the PRESENT S-box and up to 48\% for the merged DES S-box circuit. The area savings from our approach generally increases with the size of the circuit. The modest incremental cost of going from 8 to 16 PRESENT S-boxes is due to the limited size of the circuit. Note that our savings are conservative, as the area cost of the randomly generated solutions do not include the additional costs that would be needed to stealthily connect the select inputs to supply or ground.

\begin{table}[]
	\centering
	\caption{Area comparison for merged S-box circuits}
	\setlength\tabcolsep{3 pt}
        \renewcommand{\arraystretch}{1.2} 
	\label{table:table1}
	\vspace {-5pt}
	\scalebox{0.9}{
	\begin{tabular}{lr|r|r|r|r|r|}
		\cline{3-7}
		& \multicolumn{1}{l|}{}          & \multicolumn{2}{l|}{Random}                          & \multicolumn{1}{l|}{\multirow{2}{*}{GA}} & \multicolumn{1}{l|}{\multirow{2}{*}{GA+TM}} & \multicolumn{1}{l|}{\multirow{2}{*}{Improvement(\%)}} \\ \cline{1-4}
		\multicolumn{1}{|l|}{Circuit}                  & \multicolumn{1}{l|}{\#S-boxes} & \multicolumn{1}{l|}{avg} & \multicolumn{1}{l|}{best} & \multicolumn{1}{l|}{}                    & \multicolumn{1}{l|}{}                       & \multicolumn{1}{l|}{}                                 \\ \hline
		\multicolumn{1}{|l|}{\multirow{4}{*}{PRESENT}} & 2                              & 54                       & 42                        & 41                                       & 39                                          & 7                                                     \\ \cline{2-7} 
		\multicolumn{1}{|l|}{}                         & 4                              & 108                      & 84                        & 74                                       & 65                                          & 23                                                    \\ \cline{2-7} 
		\multicolumn{1}{|l|}{}                         & 8                              & 205                      & 164                       & 118                                      & 101                                         & 38                                                    \\ \cline{2-7} 
		\multicolumn{1}{|l|}{}                         & 16                             & 248                      & 213                       & 183                                      & 141                                         & 34                                                    \\ \hline
		\multicolumn{1}{|l|}{\multirow{3}{*}{DES}}     & 2                              & 257                      & 217                       & 200                                      & 195                                         & 10                                                    \\ \cline{2-7} 
		\multicolumn{1}{|l|}{}                         & 4                              & 496                      & 447                       & 257                                      & 242                                         & 46                                                    \\ \cline{2-7} 
		\multicolumn{1}{|l|}{}                         & 8                              & 923                      & 805                       & 473                                      & 416                                         & 48                                                    \\ \hline
	\end{tabular}
	}
\end{table}

%% file: conclusion.tex
In this paper we have proposed an automation technique for designing circuits that can plausibly implement a number of chosen functions. Our procedure comprises synthesis and optimization of pin assignments to maximize shared logic between the functions, and a technology mapping step that deploys camouflaged cells while ensuring that all desired functions are plausible in the final circuit. For the problem of S-box design, our technique saves up to 38\% area in PRESENT S-boxes and 48\% in DES S-boxes. This approach can find wide application in a number of practical scenarios where the adversary has partial information about what functions would be viable in an obfuscated design.
\par
\textbf{Acknowledgement:} This work has been supported by a grant from the National Science
Foundation (NSF) under award CNS-1563829 and by University of Massachusetts, Amherst.
